\documentstyle[preprint,aps,pra]{revtex}
\begin{document}
\draft

\newcommand\beq{\begin{equation}}
\newcommand\eeq{\end{equation}}
\newcommand\bea{\begin{eqnarray}}
\newcommand\eea{\end{eqnarray}}
\newcommand\gt{\tilde{g}}
\tightenlines
\title{Dilute Bose gas in a quasi two-dimensional trap}
\author{ Brandon P. van Zyl, R.K. Bhaduri, and Justin Sigetich} 
\address{ Department of Physics and Astronomy,\\McMaster University,
Hamilton, Ontario,\\ Canada L8S 4M1}
\maketitle

\begin{abstract}
We investigate the behavior of a dilute quasi two-dimensional, 
harmonically confined,
weakly interacting Bose gas within the finite-temperature Thomas-Fermi 
approximation.  We find that the thermodynamic properties of the system are
markedly different for repulsive and attractive interactions.  Specifically, in
contrast to the repulsive case, there appears to be a phase transition when 
the atoms interact with an attractive pseudo-potential, in the sense that 
there is no self-consistent solution for the normal ground state below a 
certain temperature $T^{\star}$.  These numerical findings are supported 
by analytical investigations of the thermodynamics of the system in the 
complex fugacity plane, and within the random-phase approximation. 
We also show that the temperature $T^{\star}$ can be interpreted as the
limiting temperature below which the system cannot be described as a
collection of noninteracting {\em haldons}.

\vskip .5 true cm

\end{abstract}

\section{Introduction}

Over the last decade or so, there have been a number of theoretical 
studies of Bose-Einstein
condensation (BEC) in inhomogeneous low-dimensional Bose systems\cite{bagnato,shevchenko,ketterle,mullin,pitaevskii,haugset,bhaduri,petrov,adhikari}. 
In principle, such low-dimensional systems can be
created in the laboratory through a suitable manipulation of the potential used to
trap the ultra-cold atoms.  In this paper, we will be primarily interested in 
the situation
for which the confinement in the $z$-direction is much steeper than the
in-plane confinement, so that the system can be viewed as being (quasi) 
two-dimensional (2D) in the sense that the motion of the atoms in the 
$z$-direction is ``frozen out''. 

It is well-known that a homogeneous 2D Bose
gas has no long range order (at $T\neq 0$), and so a BEC transition at any 
(non-zero) finite temperature cannot occur.
However, in a 2D harmonically confined system with oscillator frequency
$\omega_0$, it is easily shown that an ideal Bose gas has a critical BEC 
temperature 
\beq
T_c^{(0)}=(6/\pi^2)^{1/2}N^{1/2}\hbar\omega_0, 
\label{ideal}
\eeq
which, for $N\gg 1$, is much larger than the oscillator gap $\hbar\omega_0$. 
It is interesting to consider the limit
$N\rightarrow \infty$, $\omega_0\rightarrow 0$, such that
$N^{1/2}\omega_0 = $ constant.  In this situation, 
the critical BEC transition temperature $T_c^{(0)}$ remains the same.  
Note that this 
{\em is not} the usual thermodynamic limit, which in 2D would 
demand 
that $N\omega_0=$ constant, resulting in no BEC. We have previously 
shown that a self-consistent Thomas-Fermi (SCTF) model\cite{bhaduri}
(i.e., with no condensate density), has a solution at all temperatures
provided the mean-field contribution of a repulsive zero range interaction
is included, no matter how weak.
This result contradicts a previous claim in the literature\cite{mullin} that 
no self-consistent solution of the TF equation in a trap is
possible below a certain critical temperature $T_c$.  
(It is worth pointing out
that in Ref.~\cite{mullin}, the absence of a solution below $T_c$
was interpereted as a phase transition in the sytem, 
but {\em not} a phase transition to the BEC state.)
The absence of a phase transition for the repulsive case was also 
confirmed analytically by examining the branch-points of the density in the complex 
fugacity plane. 
Note, however, that the SCTF method cannot address the important issue 
of phase fluctuations in the condensate\cite{petrov}.
It was also found that the such a system could be described by 
noninteracting {\it haldons} obeying the generalized fractional exclusion 
statistics (FES)\cite{haldane}.
This has been recently confirmed by Hansson {\it et al.}~\cite{hansson}.

For a dilute Bose gas in a three dimensional trap, where the average distance 
between the atoms is much larger than the scattering length, the 
Gross-Pitaevskii (GP) approach\cite{gp} of using a zero-range pseudo-potential 
with a strength given by $4\pi\hbar^2 a/m$, (where $a$ is the s-wave 
scattering length, and m is the mass of the atom) is very successful. For 
a strictly 2D system however, the strength has a logarithmic 
energy dependence\cite{shevchenko}.  Recently, Petrov 
{\it et al.}~\cite{petrov} have deduced the strength of the quasi-2D system  
by considering free two-particle scattering in the plane of the gas, 
and strong harmonic confinement in the $z-$direction (i.e., $\omega_0 \ll 
\omega_z$). 
The effective two dimensional 
pseudo-potential may be written as a zero-range momentum-dependent 
interaction with a strength given by~\cite{note}
\beq
{2\pi\hbar^2\over m}g = {2\sqrt{2\pi}\hbar^2\over m}[\ell_z/a - 
{1\over {\sqrt{2\pi}}}\ln(\pi p_{12}^2 \ell_z^2)]^{-1}~,
\label{pseudo}
\eeq
where $\bf{p_{12}}=({\bf{p}}_1-{\bf{p}}_2)/2$ is the relative momentum between 
two bosons. 
In the above equation, $m$ is the mass of the atom, $a$ the s-wave 
scattering length, and $\ell_z = \sqrt{\hbar/m\omega_z}$. Note that 
the kinetic energy of relative motion, $E=p_{12}^2/m$, reduces to 
twice the single-particle energy in the frame where the CM momentum 
${\bf{P}}=({\bf{p}}_1+{\bf{p}}_2)/2 =0$. To simplify actual calculations 
in the many-body problem with 
this coupling strength, the momentum-dependence in Eq.(\ref{pseudo}) is 
replaced~\cite{petrov} by a density dependence by putting 
$p_{12}^2=m E=2m|\mu|$, where $\mu$ is the chemical potential. For the bosons 
in a harmonic oscillator trap, the spatial density is inhomogeneous, and the 
local chemical potential $\mu(r)=\mu-{1\over 2}m\omega_0^2r^2$ will be used. 
For the situation when $\ell_z\gg a$, the first term on the right-hand 
side (RHS) of
Eq.~(\ref{pseudo}) dominates over the second, and the logarithmic term may be
neglected. This yields a ``constant'' interaction strength,
\beq
{2\pi\hbar^2\over m}\tilde{g} = {2\sqrt{2\pi}\hbar^2\over m}a/\ell_z~,
\label{bold}
\eeq
which agrees with our definition for the coupling constant in an earlier
work\cite{bhaduri}.

In this paper, we 
perform finite-temperature SCTF calculations with the above 
density dependent 
strength factor in the pseudo-potential. Both the repulsive, and the attractive 
cases are considered for realistic values of the physical parameters. 
For the repulsive case ($g>0$), the earlier conclusion that there is no BEC at 
any temperature remains unaltered. For the attractive case ($g<0$) on the 
other hand, we find compelling numerical and analytical evidence suggesting the 
existence of a phase transition at a $T^{\star}$ which is 
greater than the critical temperature for BEC given by 
Eq.~(\ref{ideal}). 
The outline for the rest of the paper is as follows. In
Sec.~II, we present numerical results of the SCTF calculations 
for both $\tilde{g}$ and $g$ as defined above, as well as some
analytical arguments explaining the origin of the critical temperature
$T^{\star}$.  
In Sec.~III we provide additional support for our numerical results by 
analytically investigating the thermodynamic properties of the system at
zero and finite-temperatures.  
Finally, in Sec.~IV, we present our concluding remarks.

\section{Thomas-Fermi Calculation}
\subsection{Repulsive Interaction}
We consider bosonic atoms in a two-dimensional harmonic oscillator trap 
above the critical temperature, if any. In this subsection, a repulsive 
interaction is used, while in the next subsection the interaction is taken to 
be attractive. In the SCTF method\cite{bhaduri},
the inhomogeneous number density of the interacting bosons is given by 
\beq
n(r)=\int {d^2p/(2\pi\hbar^2)\over {[\exp[({p^2\over {2m}}+{1\over 2}m\omega_0^2
r^2+U(n(r))-\mu)\beta ]-1 ]}}~,
\label{tempTF}
\eeq
where $U(n(r))={2\pi\hbar^2\over m} \tilde{g}n(r)$ is the mean-field 
potential\cite{exch} generated by the zero-range two-body pseudo-potential with 
the constant 
strength $\tilde{g}$, and $\mu$ is the chemical potential that serves to fix
the total number of atoms in the trap, 
\beq
N = \int n({\bf r})~d^2{\bf r}.
\label{totalN}
\eeq
To obtain $n(r)$ for a fixed $N$, Eqs.~(\ref{tempTF}) and (\ref{totalN})
have to be solved self-consistently.
(In our numerical work, we have scaled all lengths by the
characteristic in-plane oscillator length $\ell_0 = \sqrt{\hbar/m\omega_0}$,
and all energies by $\hbar\omega_0$.)  The case for $\tilde{g} > 0$, 
has already been addressed in an earlier paper\cite{bhaduri}, where it 
was shown 
that the self-consistent solution for $n(r)$ could be obtained for any 
$T\geq 0$, indicating that there is no BEC. We first want to check if the 
same conclusion holds when $\tilde{g}$ is replaced by the more realistic
momentum dependent $g$, as given by Eq.~(\ref{pseudo}).
As noted before, we replace the $p_{12}-$dependence in 
$g$ by an effective density dependence. Under this assumption, the 
$p-$integration in (\ref{tempTF}) can be done analytically. At 
zero-temperature, the result is given by   
\beq
{2\pi \hbar^2\over m} g n(r)=(\mu-{1\over 2}m\omega_0^2 r^2)~.
\label{zeroT}
\eeq
The RHS of Eq.~(\ref{zeroT}) is precisely the local chemical potential 
$\mu(r)$. Hence we 
replace the $p_{12}^2$ on the RHS of Eq.~(\ref{pseudo}) by 
$(2\pi\hbar)^2 g n(r)$ 
and $g$ itself is determined self-consistently by Eqs.~(\ref{totalN}) and 
(\ref{zeroT}).  At finite-temperatures, we continue 
to follow the same procedure, but in this case $g(r)$ and $n(r)$ must be 
found self-consistently from Eqs.~(\ref{tempTF}) and 
(\ref{totalN}).  Note that Petrov {\it et al.} \cite{petrov} in one of their 
footnotes recommend an identical prescription. 
Following this method, we still 
find that for a fixed boson number $N$, Eqs.~(\ref{tempTF}) and (\ref{totalN})
may be 
solved self-consistently for all $T\geq 0$, indicating that there is no 
phase transition for a repulsive interaction. 
The behavior of the chemical potential $\mu$ as a 
function of $T/T_c^{(0)}$ is shown in Fig. 1.  
The lowest solid curve, for 
noninteracting bosons, shows the usual discontinuity in $\mu$ at 
$T/T_c^{(0)}=1$.  The next higher pair of closely spaced 
curves show the results for $N=10^4$ Rb$^{87}$ atoms in a trap with 
with $\hbar\omega_0 = 0.6$ nK, $\omega_z=10^3 \omega_0$, and $a=5.8$ nm. 
The solid curve 
shows the result for $\tilde{g}=0.05$, obtained from the above physical 
parameters, while the slightly lower dashed curve is for the corresponding  
$g$ given by Eq.~(\ref{pseudo}). We note that the difference between the two 
is negligible for the parameters chosen above, which suggests that FES is
still obeyed to a good approximation for relatively small values of $g$. 
In the same figure, the top 
three curves are for an artificially boosted interaction, with the 
continuous curve given by $\tilde{g}=0.25$. Note that this could in 
principle be achieved in a variety of ways by altering both $\ell_z$ and $a$. 
Two limiting  cases are given by keeping $\ell_z$ constant and decreasing 
$a$ by a factor of five, or keeping $a$ fixed and increasing $\ell_z$ by a factor 
of five. Although both procedures yield the same $\tilde{g}$, we see from 
Eq.~(\ref{pseudo}) that $g$ gets affected more when $\ell_z$ alone is altered. 
In Fig. 1, the lowest (dashed) curve is for this case, while the next higher 
one is for the case where only $a$ is decreased by a factor of five. We 
see that in all cases with a repulsive interaction, the main conclusion that 
there is no strict BEC continues to hold.  

Even though there is no BEC in the sense that the SCTF solution is obtained 
without paying any special attention to the lowest single-particle quantum 
state, it is found that there is a significant enhancement in the peak of the 
density distribution as the temperature is reduced below $T_c^{(0)}$, the 
critical temperature for the noninteracting case. This is shown in Fig. 2 for 
$\tilde{g}=0.25$ (continuous curves) at $T/T_c^{(0)}=0.8$ and $1.1$. When the 
density-dependent version of the coupling $g$ is used, the enhancement is 
even more pronounced. The reason for this behavior is clear from the inset, 
where the effective $g(r)$, as obtained from the self-consistent solution, 
is shown as a function of $r$. We note that for the density-dependent 
case, the interaction $g(r)$ is considerably  weaker than the constant 
$\tilde{g}$, thus causing less suppression of the central density.  
%%%%%%%%%%%%%%%%%%
%%%%%%%%%%%%%%%%%%
\subsection{Attractive Interaction}
We first take the example of Li$^7$ atoms for which the s-wave scattering 
length is $a=-1.45$ nm. For $\hbar \omega_0 = 7$ nK 
(the corresponding oscillator 
angular frequency $\omega_0$ in the plane is $2 \pi \times 145$ rads/s )
and $\omega_z = 10^3 \omega_0$,
the coupling constant $\tilde{g}=-0.012$. On the other hand, the interaction 
strength could be considerably larger for Cs$^{135}$ atoms, which have 
$a=-60$ nm. In Fig. 3, we display for $N=10^4$ atoms of Li$^7$ the 
behavior of the chemical 
potential $\mu$ as a function of $T$ obtained self-consistently. No solution 
is found for $T\leq T^{\star}$, irrespective of whether one uses 
${\tilde{g}}$ or $g$. The highest lying solid curve is for $\tilde{g}=-0.012$, 
and 
difference between this and the density-dependent form of $g$ for this weak 
interaction is too small to be shown in the same scale. In the same figure, 
we also display the effect of boosting the attractive interaction strength 
nearly ten-fold to $\tilde{g}=-0.1$. Not only does the temperature 
$T^{\star}$ increases significantly, but it is also possible now to differentiate 
the curves for the cases where the density-dependent form of $g$ is used rather than 
$\tilde{g}$. Furthermore, when the boosting of the coupling $g$ is done by 
altering $\ell_z$ rather than $a$, $T^{\star}$ is larger.
Nevertheless, the behavior of the chemical potential with temperature, for  
$T>T^{\star}$ does not alter significantly with the strength of the attractive 
interaction.  Comparing Fig.~3
with the corresponding repulsive cases (see Fig.~1), we note that $\mu$ showed 
a much larger variation, but only in the range $T\leq T_c^{(0)}$, which is 
in any case inaccessible for the attractive interactions.

Some insight as to why no self-consistent solution 
for attractive interaction strength $\tilde{g}$ may be found by performing 
the $p-$integration in Eq.~(\ref{tempTF}) analytically 
(valid for either sign of $\tilde{g}$ ).  The result is given by
\beq 
(\mu-{1\over 2}m\omega_0^2 r^2)~={2\pi \hbar^2\over m} \tilde{g} n(r) 
+ {1\over \beta}~\ln [1-\exp(-2\pi\hbar^2 n(r) \beta/m)]~.
\label{true}
\eeq
For a negative 
$\tilde{g}$, both terms on the RHS are negative, and therefore $\mu$, which 
is independent of $r$, must be negative at all temperatures. For this case,  
if we take the above expression literally at $T=0$, the density is 
given by 
\beq
n(r)= {m\over {2\pi|\tilde{g}| \hbar^2}} (|\mu|+{1\over 2}m\omega_0^2 r^2)~.
\eeq
This solution for $n(r)$ is rejected since it keeps increasing with $r$ ! 
We demand a normalized density $n(r)$ that is a monotonically decreasing 
function of $r$, which is only found for $T>T^{\star}$. 
By differentiating Eq.~(\ref{true}) with respect to $r$, we find that for 
$dn/dr \leq 0$ for all values of $r$, we must have  
\beq
[\exp ({2\pi\hbar^2\over m}n(r)\beta)-1]^{-1} - |\tilde{g}| \geq 0~.
\eeq
This may be satisfied only for $T\geq T^{{\star}}$, which is related to the central 
density at this temperature $n(0)$ by the relation 
\beq
T^{{\star}}={{2\pi\hbar^2\over m} n(0)\over {\ln (1+1/|\tilde{g}|)}}~.
\label{new}
\eeq

It is interesting that the above equation may also be derived using the 
the generalized exclusion statistics of Haldane\cite{haldane}. 
It has been shown~\cite{bhaduri,hansson} that when the interaction strength 
of the zero-range pseudo-potential is taken to be a constant (which we denote 
by $\tilde{g}$ in this paper), the interacting boson system may be mapped on 
to a noninteracting system of haldons, whose statistical 
occupancy factor of a 
single particle state $i$ with energy $\epsilon_i$ is given 
by 
\beq
\eta_i=(w_i+\tilde{g})^{-1}~, 
\label{ocu}
\eeq
which must be $\geq 0$. Note that $w_i$ obeys the relation 
$$w_i^{\tilde{g}}(1+w_i)^{1-\tilde{g}}=\exp[(\epsilon_i-\mu)\beta]~.$$  
In what follows, we go to the limit 
$\omega_0\rightarrow 0$, $N\rightarrow \infty$, such that 
$N^{1/2}\omega_0 = $ constant. We then identify the central 
density $n(0)$ of the harmonic oscillator potential as the uniform density 
in this limit. The number density $n(0)$ of these haldons at the center of the 
harmonic oscillator potential is given by~\cite{bhaduri} 
\beq
n(0)={m\over {2\pi\hbar^2\beta}}~\ln {1+w_0\over w_0}~,
\label {aya}
\eeq
where $w_0$ defines the occupancy of the lowest quantum state through 
Eq.~(\ref{ocu}). We now use these 
relations for the attractive interaction strength $\tilde{g}=-|\tilde{g}|$. 
Note from Eq.~(\ref{ocu}) that for the occupancy $\eta_0$ to be positive 
definite, $w_0\geq |\tilde{g}|$, and that the occupancy tends to infinity when 
$w_0=|\tilde{g}|$. Substituting this limiting value of $w_0$ in 
Eq.~(\ref{aya}) at $T^{\star}$,  we immediately obtain Eq.~(\ref{new}). 
We should regard $T^{\star}$ as the limiting temperature below which the 
system cannot be described as a collection of ideal haldons.

In Fig. 4, the self-consistent spatial density $n(r)$ is shown for 
$N=10^4$ Li$^7$ atoms in the quasi 2D trap, with 
$\tilde{g}=-0.012$. For this very weak interaction, the corresponding curve 
for the density-dependent coupling strength $g$ cannot be differentiated on 
this scale. We see that not only is there there a marked enhancement in the 
central density when $T$ is decreased from $1.1$ $T^{\star}$ to $T^{\star}$, 
but the 
shape of the density distribution is also different. We cannot, of course, 
get any solution for $T <T^{\star}$. It may be instructive to find a relation 
between $T^{\star}$ and $N$ analogous to Eq.~(\ref{ideal}), which will of 
course 
also depend on the interaction strength $\tilde{g}$. Empirically, we find 
that this is given by 
\beq
T^{\star}=\gamma(|\tilde{g}|) 
\left(\frac{N}{\zeta(2)}\right)^{1/2}\hbar\omega_0~,
\label{emp}
\eeq 
where $\gamma(|\tilde{g}|)=[1+{6\over 5}(|\tilde{g}|)^{2/3}]$.
This relation is quite accurate even for small $N$ and large $\tilde{g}$, 
as depicted in Fig. 5. In this figure, we show the numerically calculated 
points by circles, and the continuous curves are derived from 
Eq.~(\ref{emp}).  Although the solid
curves fitting the data have been extended down to $N=0$ at $T=0$, 
we were not able to numerically access stable values of 
$N$ for $T^{\star} \lesssim 8$,
and so Eq.~(\ref{emp}) is only used to fit the data down to a {\em finite}
temperature $T^{\star}\gtrsim 8$.
Also note from Eqs.~(\ref{ideal}) and (\ref{emp}) that 
$T^{\star}/T_c^{(0)}$ is independent of 
$N$, and therefore we can draw a ``universal'' curve of $\tilde{g}$ versus 
$T^{\star}/T_c^{(0)}$, which is shown by the inset in Fig.~5. 
The normal bosonic 
phase is above the continuous curve shown in this inset, but the SCTF method 
used cannot access the region below.

\section{Analytical considerations}
In this section, we wish to address the question that naturally arises
from the numerical work presented in Sec. IIB; namely, are our numerical
calculations suggesting that there is phase transition in the quasi-2D trap for
attractive interactions ?
Since we are unable to answer this question within the framework 
of the SCTF model, we now turn our attention to various analytical 
approaches which may aid in our assessment of the numerical results 
presented above.
%%%%%%%%%%%%%%%
\subsection{Zero-Temperature Variational Calculation}
Let us consider the 2D analogue of the GP energy functional, namely
\beq
E=\int d^2r~\left[ {\hbar^2\over {2m}} |\nabla \psi|^2 + {m\over 2} r^2 
\omega_0^2 |\psi|^2 + {1\over 2} {2\pi\hbar^2\over m} \tilde{g} |\psi|^4 
\right ]~,
\label{gp}
\eeq
where $\tilde{g}$ is given by Eq.~(\ref{bold}), and may be of either sign. 
In Eq.~(\ref{gp}), the function  $\psi(r)$ represents the wave function
of a system for which there is macroscopic occupation of the lowest quantum
state (i.e., all $N$ bosons are in the ground state): 
\beq
\psi(r)={1\over b}\sqrt{N\over \pi} \exp(-{r^2\over b^2})~,
\eeq
where $b$ is taken to be a variational parameter.  We find that the energy 
is a minimum for 
\beq
b=\sqrt{\hbar\over {m\omega_0}} (1+N\tilde{g})^{1/4} ~.
\label{bmax}
\eeq
For an attractive interaction, $\tilde{g}=-|\tilde{g}|$, Eq.~(\ref{bmax})
implies that the condensate collapses for $$N|\tilde{g}|\geq 1.$$
This rough estimate is in good agreement with the 
{\em numerical}
estimate of $N|\tilde{g}| \geq 0.94$ by Adhikari~\cite{adhikari} 
which was carefully obtained through a self-consistent solution of the 
2D GP equations.  For a weakly interacting gas like 
Li$^7$ with $\tilde{g}=-0.012$, we find that 
the condensate is unstable for $N >83$. 
%%%%%%%%%%%%%%%%%%%%%%%%%%%5
\subsection{Finite-Temperature Thermodynamics}
This subsection, deals with the study of the behavior of the branch points 
of the quantity $u=2\pi\hbar^2\beta n(r)/m$ in the complex fugacity plane 
$Z=\exp(\beta\mu)$ as a 
function of the coupling constant $\tilde{g}$.  This is analogous to the 
analysis that was made by Sutherland while studying the thermodynamics of a  
one-dimensional gas interacting with an inverse square 
potential\cite{sutherland}.  The important point to be taken from
Sutherland's work is that
a branch point of $u$ on the positive real axis of $Z$ indicates a phase 
transition in the system.

We begin by noting that Eq.~(\ref{true}) can be rewritten as
\beq
Z~ e^{-\beta m \omega_0^2 r^2/2}=2~e^{(\tilde{g}-1/2)} \sinh {u\over 2}~.
\label{true_2}
\eeq 
As before, going to the limit $N\rightarrow \infty$, 
$\omega_0 \rightarrow 0$ with $N^{1/2}\omega_0 =$ constant, Eq.~(\ref{true_2})
reduces to the $r$-independent relation 
\beq
Z=e^{\tilde{g}u}-e^{(\tilde{g}-1)u}. 
\label{complex}
\eeq
We note that for $\tilde{g}=0$, the above equation gives $u=-\ln (1-Z)$, 
implying that $u$ has a branch point at $Z=1$, i.e. for $\mu=0$. This is 
precisely where BEC takes place for a noninteracting gas in a harmonic trap. 
For $\tilde{g}=1/2$, the branch points of $u$ are at $Z=\pm 2i$, and for 
$\tilde{g}=1$, the branch point is at $Z=-1$. 
For arbitrary values of $\tilde{g}$, the branch points of $u$ can be obtained
by examining the analytical structure of the solutions to the equation
($x \equiv \exp(u)$)
\beq
x^{\tilde g} - x^{\tilde{g} - 1} - Z = 0.
\label{complex_2}
\eeq
Following Sutherland's prescription\cite{sutherland}, we obtain the
branch points of $u$, denoted by $Z_0$, for all $\tilde{g} \geq 0$:\\ \\
(i) $\underline{\tilde{g} \geq 1}$
\beq
Z_0 = -\frac{1}{\sqrt{\tilde{g}(\tilde{g}-1)}}
\left(\frac{\tilde{g}-1}{\tilde{g}}\right)^{(\tilde{g}-1/2)}~,
\label{Z01}
\eeq
\\
(ii) $\underline{1 > \tilde{g} \geq 0}$
\beq
Z_0 = e^{\pm\pi i \tilde{g}}\frac{1}{\sqrt{\tilde{g}(1-\tilde{g})}}
\left( \frac{1-\tilde{g}}{\tilde{g}}\right)^{(\tilde{g}-1/2)}~.
\label{Z02}
\eeq
In Fig. 6, the loop and the 
line along the negative real axis $-1\leq Z \leq 0$ in the 
complex fugacity plane shows the branch points of $u$ for positive values 
of $\tilde{g}$ given by Eqs.~(\ref{Z01}) and (\ref{Z02}). 
Since there is no branch point on the real fugacity axis for 
$\tilde{g}>0$, there is no phase transition, in agreement with our numerical 
calculations. 

The situation for $\tilde{g} < 0$ can be addressed by either solving 
Eq.~(\ref{complex_2}) directly, or by noticing that the 
transformation~\cite{niall} 
\beq
\tilde{g}\rightarrow -\tilde{g}+1,~~~~~u\rightarrow -u,~~~~~Z\rightarrow -Z~,
\label{transf}
\eeq
leaves Eq.~(\ref{complex}) unchanged.  More specifically, under
Eq.~(\ref{transf}), the 
branch points of $u$ on the negative $Z-$axis between $Z_0 \in [-1,0]$ 
for $1\leq\tilde{g}< \infty$ are now shifted to the positive side of the 
axis, namely, $Z_0\in [0,1]$ for all $-\infty < \tilde{g}\leq 0$. 
The branch points for $0<\tilde{g}<1$ are all mapped back onto 
Eq.~(\ref{Z02}), thereby leaving the loop structure in the complex $Z$-plane
invariant under (\ref{transf}) (see Fig.~6).
In other words, for every $\tilde{g} < 0$ we have an associated branch point on 
the {\em positive} real $Z$-axis, suggesting that there
is a phase transition in the system for attractive interactions. 
It is important to note, however, that the above analysis can offer us no
insight into the precise {\em nature} of the phase transition; is it
BEC ? Is it the Kosterlitz-Thouless transition \cite{kt} ?
Is it the collapse of the gas ?  
%%%%%%%%%%%%%%%%%%%%%%%%%%%%%%
%%%%%%%%%%%%%%%%%%%%%%%%%%%%%
\subsection{Random-Phase Approximation}
In this subsection, we will try to address the nature of the phase
transition discussed above by applying the well-known random-phase 
approximation 
(RPA)\cite{griffin} to investigate the stability of the trapped 2D Bose gas with
attractive interactions.  Our approach parallels the recent investigation by
Meuller and Baym\cite{baym}, with the main difference being that we confine
ourselves to strictly two-dimensions.  
The basic idea behind their method is to identify 
the collapse of the gas with a divergence in the density-response
function $\chi(k,\omega;\beta)$, which is a measure of the response of the
system to some external probe with wave vector $k$ and frequency $\omega$.
As in Ref.~\cite{baym}, $\chi$ is calculated
within the local-density approximation (LDA),
which amounts to replacing the response function of the inhomogeneous
system by that of a system of uniform density $n$.  In order to make contact
with the trapped gas, the uniform density $n$ is then taken to represent the
central density of the trap $n(r=0)$.

For a confined Bose gas, the
collapse is associated with an instability of the
lowest energy ``breathing'' mode of the system.  In 2D, the breathing mode
is given by
$$\delta\rho(r) \propto \left(\frac{r^2}{\ell_0^2} - 1\right)\exp(-r^2/\ell_0^2)~,$$
where $r$ is the 2D radial coordinate, and $\ell_0=\sqrt{\hbar/m\omega_0}$ 
is the characteristic
size of the cloud.  The Fourier transform of
$\delta\rho(r)$ has a maximum at $k = 2/\ell_0$, and so we look for
an instability at this wave vector.
Owing to the fact the breathing mode has a
vanishing frequency, the line of collapse is characterized by 
\beq
\chi(k=2/\ell_0,\omega=0;T^0)^{-1} = 0~,
\label{chieq0}
\eeq
where $T^0$ denotes the 
temperature for which the gas undergoes
collapse.  Note that $T^0$ need not coincide with 
the BEC transition temperature $T_c$ of the interacting gas.
Since our SCTF calculations deal only with the normal phase of the Bose gas,
we will restrict our attention to the noncondensate contribution
to $\chi(k,\omega;T)$ above, and present a more detailed analysis of
the RPA in 2D attractive Bose gases elsewhere\cite{note2}.
Within the RPA, the response function for the interacting
Bose gas has the structure~\cite{note3}
\beq
\chi^{\rm RPA}(k,\omega;T) = \frac{1}
{(1-\tilde{g}\chi_0^n)}~,
\label{rpa}
\eeq
where $\chi_0^n$ is the ``bare'' noncondensate 
response of the {\em noninteracting} system.  In 2D, we have
\beq
\chi_0^n(k,\omega;T) = \frac{1}{(2\pi)^2}\int~d^2{\bf q}
\frac{f(q-k/2)-f(q+k/2)}{\hbar\omega-(\varepsilon(q+k/2)-\varepsilon(q-k/2))}~.
\label{chin}
\eeq
In Eq.~(\ref{chin}), $f(q)$ is the Bose distribution function, and
$\varepsilon(k)=\hbar^2k^2/2m$ is the free particle kinetic energy. 
After integrating out the angular dependence, we
have
\beq
\chi_0^n(k,0;T) = -\frac{8\pi m}{(2\pi\hbar)^2 k}
\int~dqf(q)\frac{q}{\sqrt{k^2-4q^2}}~.
\label{chin_2}
\eeq

Using the above expression for $\chi_0^n$ 
we can obtain the line of collapse by setting $k=2/\ell_0$ and solving
for the roots of the denominator in Eq.~(\ref{rpa}).  
This procedure yields the
line of collapse in terms of $\mu$ and $T^0_{\rm RPA}$.   
Following the arguments made in Ref.~\cite{baym}, 
we can then relate $\mu$
to the total number of normal phase atoms, $N$, in the harmonically confined 
gas through the relation
\bea
N&=&\left(\frac{k_BT}{\hbar\omega_0}\right)^2g_2(e^{\beta\mu})~,
\label{above}
\eea
where $g_\nu(z)=\sum_pz^p/p^\nu$ is the polylogarithm function. 
Note that evaluating Eq.~(\ref{above}) at $\mu = 0$ defines the line of 
condensation for a {\em noninteracting} 2D Bose gas.  (This result can easily
be derived by integrating Eq.~(\ref{tempTF}) with $\tilde{g} = 0$ 
over all space.)

The results of our calculations for $\tilde{g} = -0.012,-0.1,-0.2$ are
shown in Fig.~7.  For comparison, we have also included as dashed lines in
the figure the
$\tilde{g} = -0.1,-0.2$ SCTF results from Sec.~IIB 
(see also Fig.~5).  This figure suggests that the line of
collapse derived from the RPA has the same functional 
relationship as the empirical formula given by Eq.~(\ref{emp}) in Sec.~IIB.  
In order to confirm this observation, we recall that the
relationship used to link
the uniform gas results to the harmonically confined gas, viz.,
Eq.~(\ref{above}), is strictly valid only for a noninteracting system.
As we saw in Sec.~IIB, the attractive interactions have the effect of
``renormalizing'' the ideal $N$ vs.~$T$ dependence
(see Eq.~(\ref{emp})), which will result
in different curves for the RPA and SCTF data.
With this in mind, we rescale  Eq.~(\ref{above}) by
$1/\gamma(|\tilde{g}|)$, and comapre the results with the SCTF
data.  After this procedure, we find that the SCTF and
RPA curves are almost indistinguishable on the scale of Fig.~7.  In this sense,
we can identify $T^0_{\rm RPA}$ with the $T^{\star}$ of Sec.~IIB.
This analysis leads us to believe
that the phase-boundary generated from the SCTF in Sec.~IIB is in fact
the line of collapse of the Bose gas as one approaches the instability from
above the BEC transition temperature $T_c$.

To close this section, we mention that we have also performed RPA 
calculations for the 
3D interacting Li$^7$ Bose gas with attractive interactions (exactly as in
Ref.~\cite{baym} but without exchange) and found that the 3D gas
is more stable than its 2D counterpart; that is, for the same 
temperature and particle number, the 3D system is always in the stable
uncollapsed phase relative to the 2D gas.
To illustrate this, we
have included as an inset to Fig.~7, the lines of collapse (solid curve)
and condensation (dashed curve) 
for the 3D system, focusing only on the $N$ and $T$ regimes that are relevant
for a comparison with the corresponding 2D system.

\section{Summary}
One of the first results of this paper was a confirmation of the fact that, 
at least within the SCTF method, there is no strict BEC
in a quasi-2D trap for atoms with repulsive interactions, even if the more 
``realistic'' momentum-dependent pseudo-potential
recently derived by Petrov {\em et  al.}~\cite{petrov}, and
given by Eq.~(\ref{pseudo}) is used.  Indeed, our results indicate that 
the much simpler density-independent coupling constant $\tilde{g}$ 
(Eq.~(\ref{bold})) is quite adequate for numerical work on 2D Bose systems.
This observation provides further support to the conclusion 
obtained earlier~\cite{bhaduri} using a {\em momentum-independent} coupling
constant $\tilde{g}$.

In the case of attractive interactions, we have found strong numerical
and analytical evidence for the existence of a phase transition in the
system.  Since the qualitative features of this phase transition are not
strongly influenced by using a more complicated momentum-dependent 
pseudo-potential, it is again sufficient to use the 
simpler $\tilde{g}$ in numerical work.  
Our numerical investigations reveal that the phase transition is 
signaled by a ``critical'' temperature
$T^{\star}>T_c^{(0)}$ below which there is no self-consistent solution
for the normal ground state of the system.  From the point of view of FES, the 
$T^{\star}$ was interpreted as the limiting temperature below which the 
system cannot be described as a collection of noninteracting haldons.  

In order to further our understanding of the numerical results, we considered
several analytical approaches.  This began in
Sec.~IIIA with a zero-temperature 2D variational calculation of the GP
energy functional.  Surprisingly, this simple analysis revealed
that if there is a BEC for attractive interactions in 2D, then the region of
stability for the condensate (i.e., prior to collapse) is severely restricted
by the condition $N|\tilde{g}|\geq 1$.  For Li$^7$, this implies 
an instability for $N > 83$.  The $T=0$ calculation was followed in
Sec.~IIIB by an investigation of the branch points of $u$ in the complex
fugacity plane.   We found that for every $\tilde{g} <0$, there is
an associated branch point of $u$ on the {\em positive} real $Z$-axis.  
This result
strongly supports our numerical calculations which indicate that there is a 
phase transition taking place in the system for attractive interactions.

Finally, in Sec.~IIIC, an RPA-type calculation was used to investigate the 
line of collapse
for a 2D Bose gas with attractive interactions.  The purpose of this
study was to try and determine the nature of the phase transition suggested
by the analysis in Sec.~IIIB.  We found that the line of instability in the
RPA calculation had the same $N$ vs.~$T$ dependence as the SCTF curves
in Sec.~IIIB.  The correspondence between the two formulations is
highly suggestive that the phase transition is in fact associated with 
the collapse of the 2D gas as one approaches the instability from above
$T_c$.  We also performed analgous RPA
calculations for the 3D Bose gas with attractive interactions, and found
that for the same $N$ and $T$, the 3D system is always in a stable,
uncollapsed phase relative to the 2D gas.  We will present a more detailed
comparison of the 2D and 3D RPA calculations elsewhere.

\section{Acknowledgments}
One of us (BVZ) would like to thank Dr. J. P. Carbotte for financial support  
as well as stimulating discussions. RKB would like to thank Dr. Niall 
Whelan for relation~(\ref{transf}). This research was supported by grants 
from the Natural Sciences and Engineering Research Council of Canada.

\begin{figure}
\caption{Chemical potential $\mu$ as a function of $T/T_c^{(0)}$.  
The parameters used to generate this figure are consistent with those of
the Rb$^{87}$ experiments.  Solid
lines are for $\tilde{g}$, dashed lines are for $g(r)$ with fixed $a$ and
variable $\ell_z$, while long-dashed-short-dashed lines are for $g(r)$ with
fixed $\ell_z$ and variable $a$.  Unless stated otherwise, this convention is 
used in all subsequent figures.}  
\end{figure}

\begin{figure}
\caption{Calculated self-consistent density profile for constant 
$\tilde{g}=0.25$
and an artificially boosted density-dependent interaction $g(r)$.  
The lowest set of curves
correspond to $T/T_c^{(0)}=1.1$ and the highest set to $T/T_c^{(0)}=0.8$.
The inset illustrates the self-consistent $g(r)$ evaluated at
$T/T_c^{(0)} = 0.8$. Notice that $g(r)$ {\em always} lies below the 
corresponding constant $\tilde{g}$ value.}  
\end{figure}

\begin{figure}
\caption{Chemical potential $\mu$ as a function of $T$ for attractive
interactions.  Open circles denote the last point for which a self-consistent
solution to Eqs.~(\ref{tempTF}) and (\ref{totalN}) could be obtained while
the arrows serve to clarify the location of $T^{\star}$.
The parameters used to generate this figure are consistent
with those of the Li$^7$ experiments.  The same line labeling convention
as in the repulsive case has been used.  See text for details.}
\end{figure}

\begin{figure}
\caption{Calculated self-consistent density profile for an attractive
interaction $\tilde{g}=-0.012$.  The lowest curve is for a temperature
$T=1.1T^{\star}$ and the highest lying curve is for a temperature
$T=T^{\star}$.  Li$^{7}$ parameters have been used.}
\end{figure}

\begin{figure}
\caption{Total number of normal phase atoms $N$ as a function of the 
critical temperature
$T^{\star}$.  From left to right, each curve is evaluated at 
$\tilde{g}=0.0,-0.1,-0.2,-1.0$.  Open circles
denote numerical data, and the continuous curves are obtained from 
Eq.~(\ref{emp}).  Inset: Universal ``phase-boundary'' separating the normal
bosonic phase from a region which is inaccessible within the SCTF 
approximation.}
\end{figure}

\begin{figure}
\caption{The thick continuous curves represent the location of branch points 
of $u$ in the
complex fugacity plane as a function of $\tilde{g}$.   Branch points of $u$
lying to the positive real $Z$-axis indicate a phase transition in the
system.}
\end{figure}

\begin{figure}
\caption{Total number of normal phase atoms as a function of temperature
$T$.  The solid curves (RPA), from left to right, are
for $\tilde{g} = -0.012, -0.1,-0.2$ and correspond to $N$ vs.~$T^0_{\rm RPA}$.
The dashed curves (SCTF),
from left to right, are for $\tilde{g} = -0.1,-0.2$ and correspond to
$N$ vs.~$T^{\star}$.  Inset: Lines of 
collapse (solid) and condensation (dashed) for the 3D Li$^7$ Bose gas 
with attractive interactions.  In this inset,
we have zoomed in to show only the $N$ and $T$ ranges relevant for a
comparison with the corresponding 2D system.  See Ref.~[19] for details.}
\end{figure}


\begin{references}
\bibitem{niall} This transformation was pointed out to us by Niall D. Whelan.
\bibitem{note2}
We have performed similar RPA calculations for $T<T_c$ which at $T=0$ are in
reasonable agreement with the simple variational calculation in Sec.~IIIA. 
Furthermore, (within the approximations of the formalism)
the RPA calculations indicate that the 2D Bose gas
with attractive interactions has a stable condensed phase 
at finite temperatures provided $N$ is below some critical 
particle number $N_c(T)$.  This is not necessarily in contradiction to
the results of Ref.~\cite{petrov} (which suggest the existence of a 
stable quasi-condensate with fluctuating phase for $T < T_c$), 
since our simple RPA formulation
cannot examine fluctuations in the phase.
\bibitem{note3}
This is the usual RPA without exchange.  To include exchange effects,
one should replace the denominator in Eq.~(\ref{rpa}) by
$(1-2\tilde{g}\chi_0^n)$.  See Ref.~\cite{griffin} for details.
\end{references}
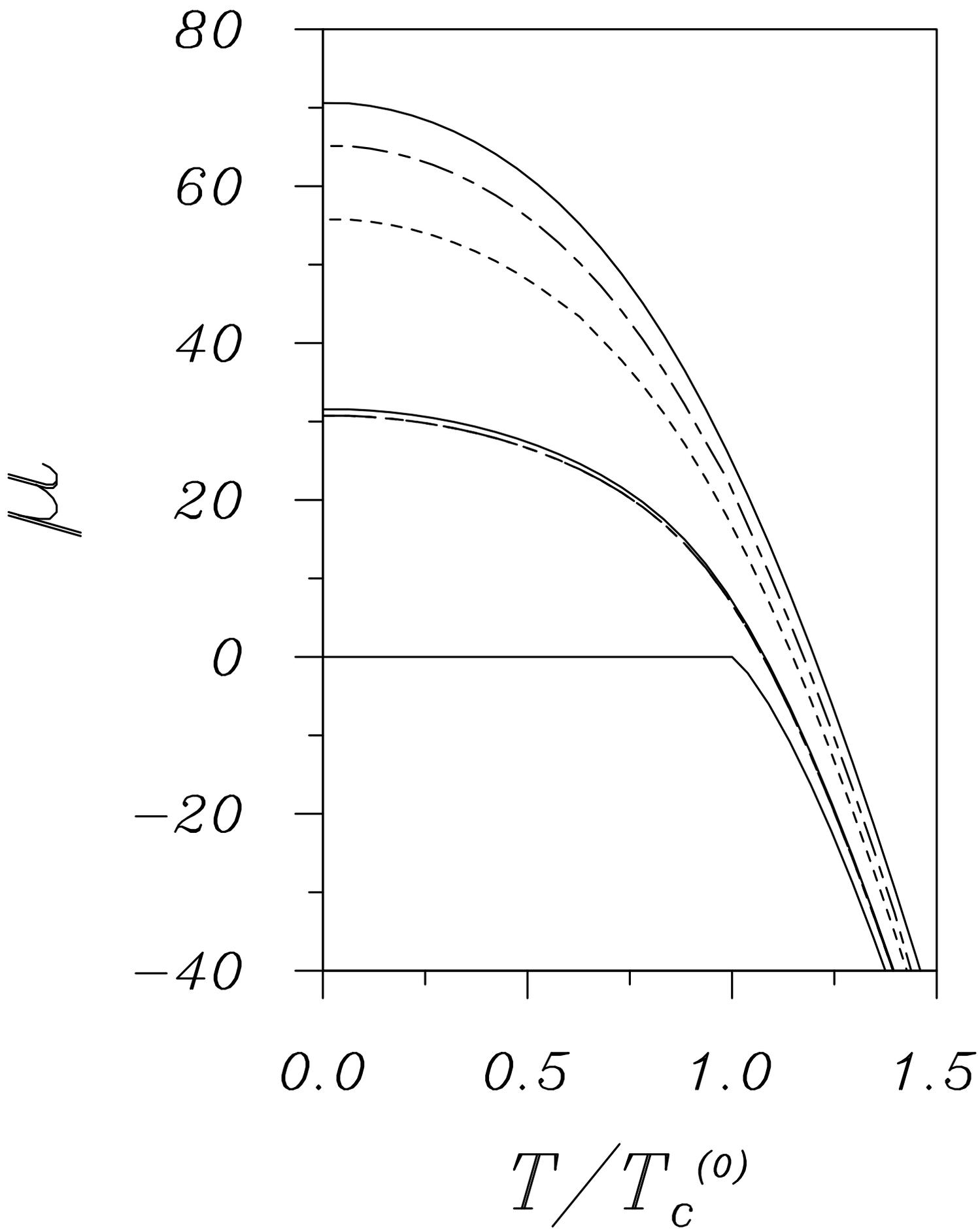
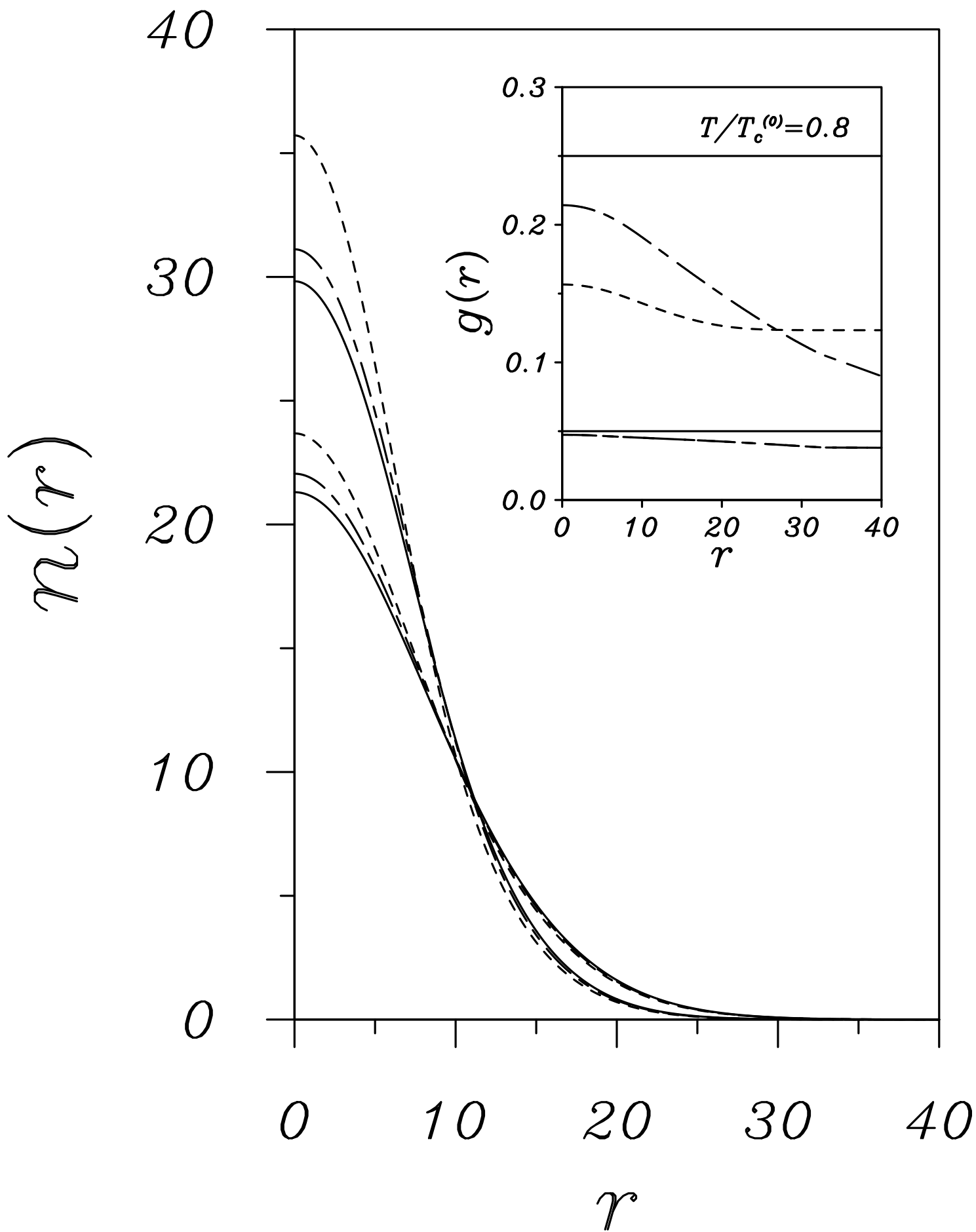
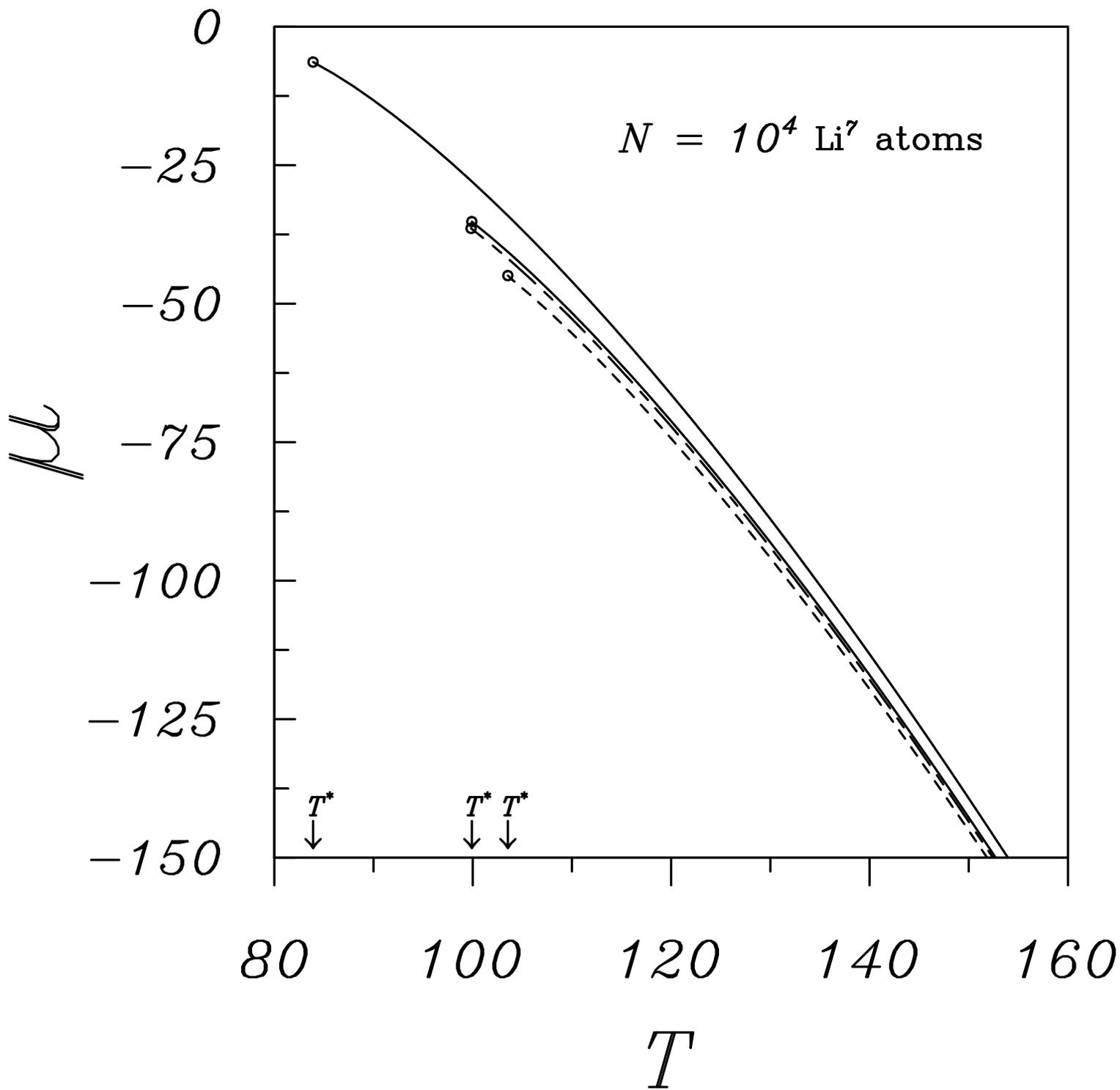
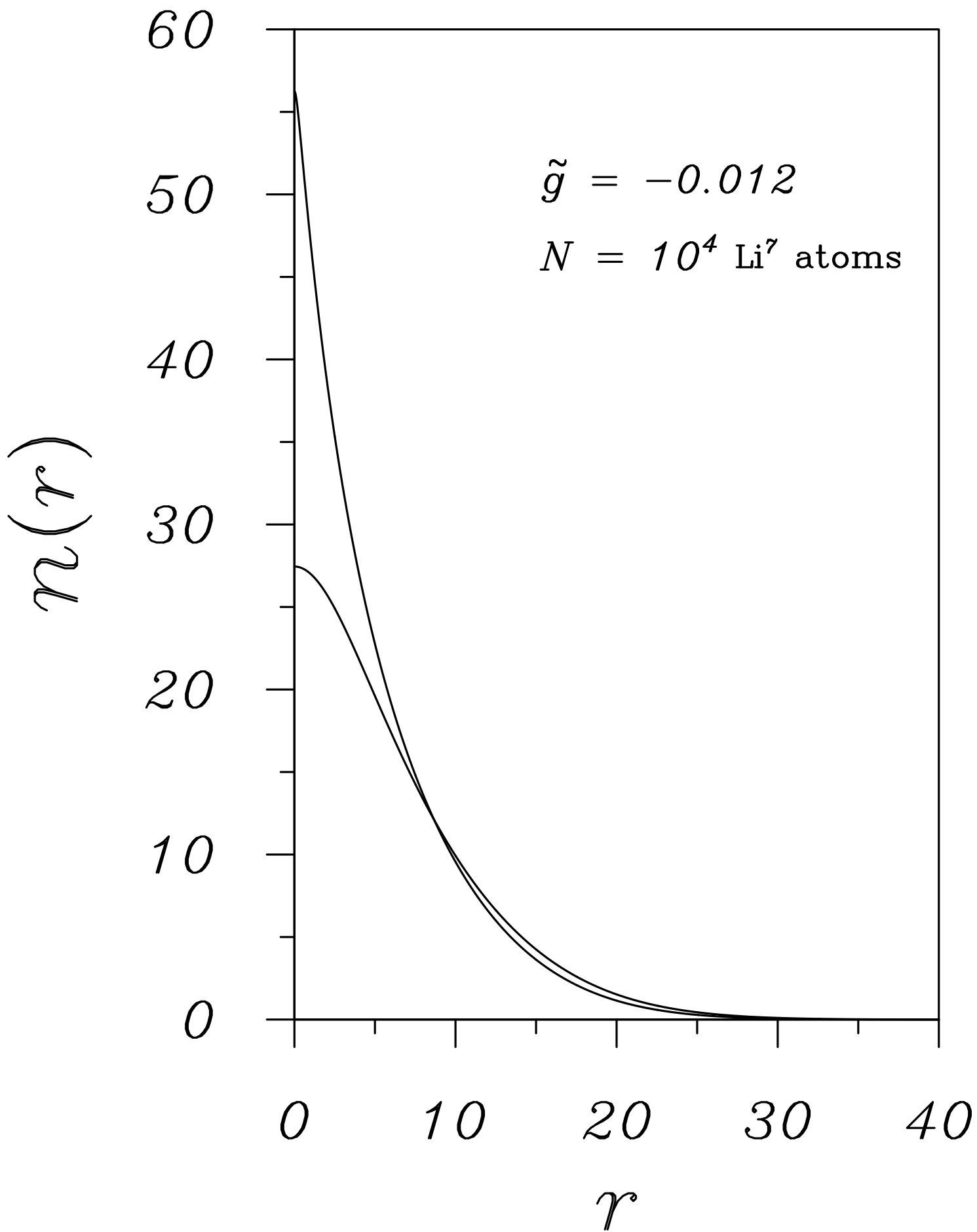
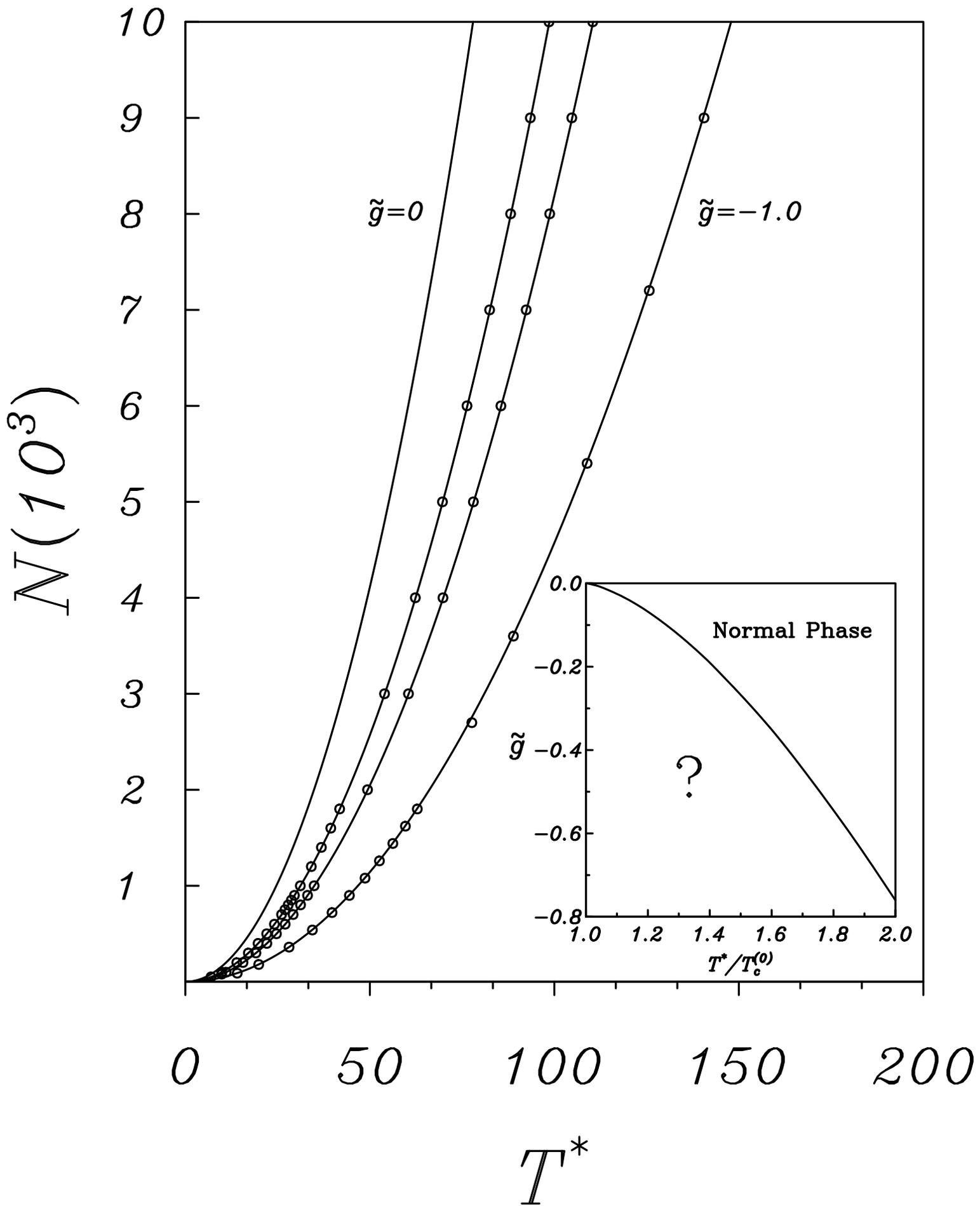
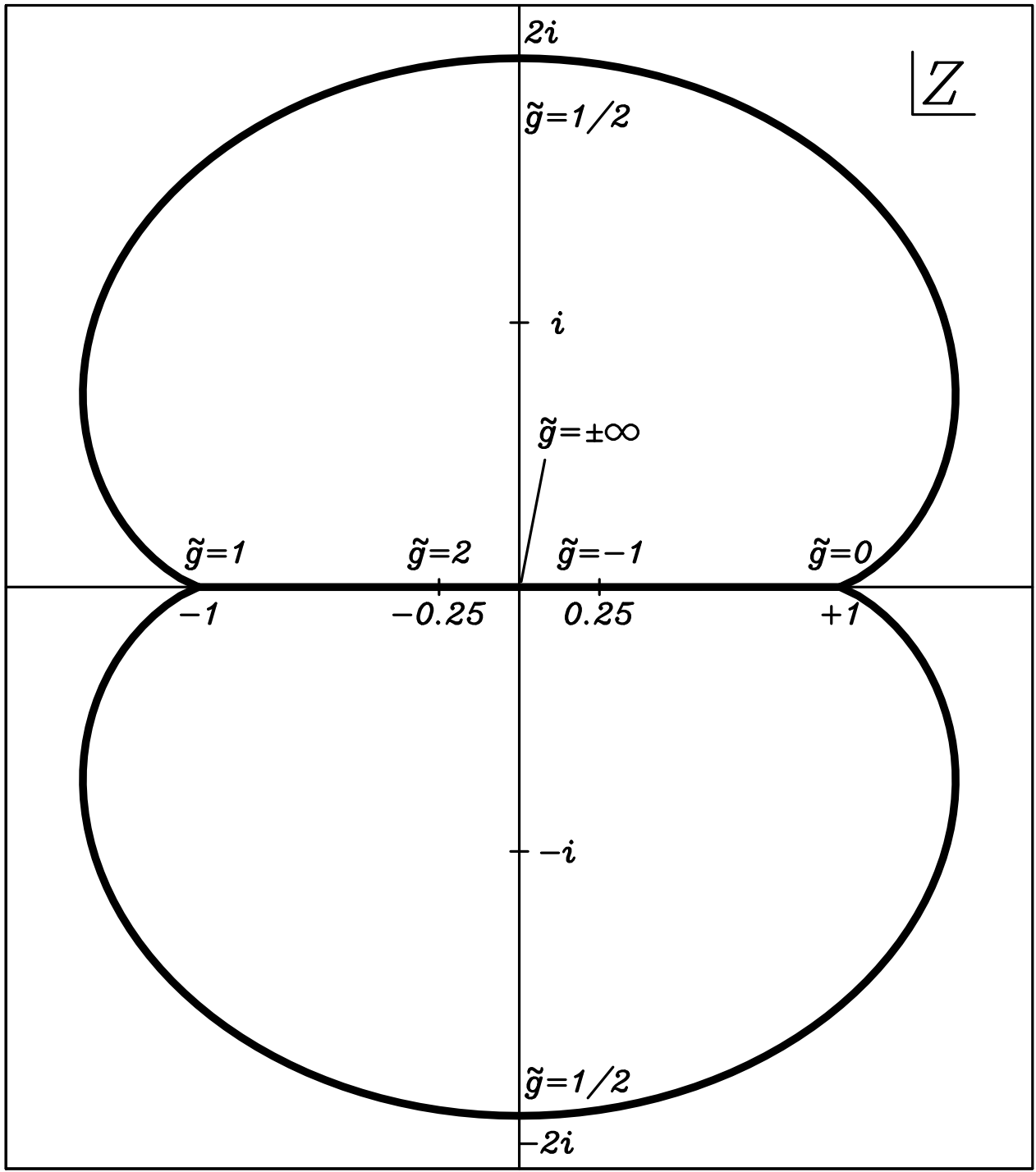
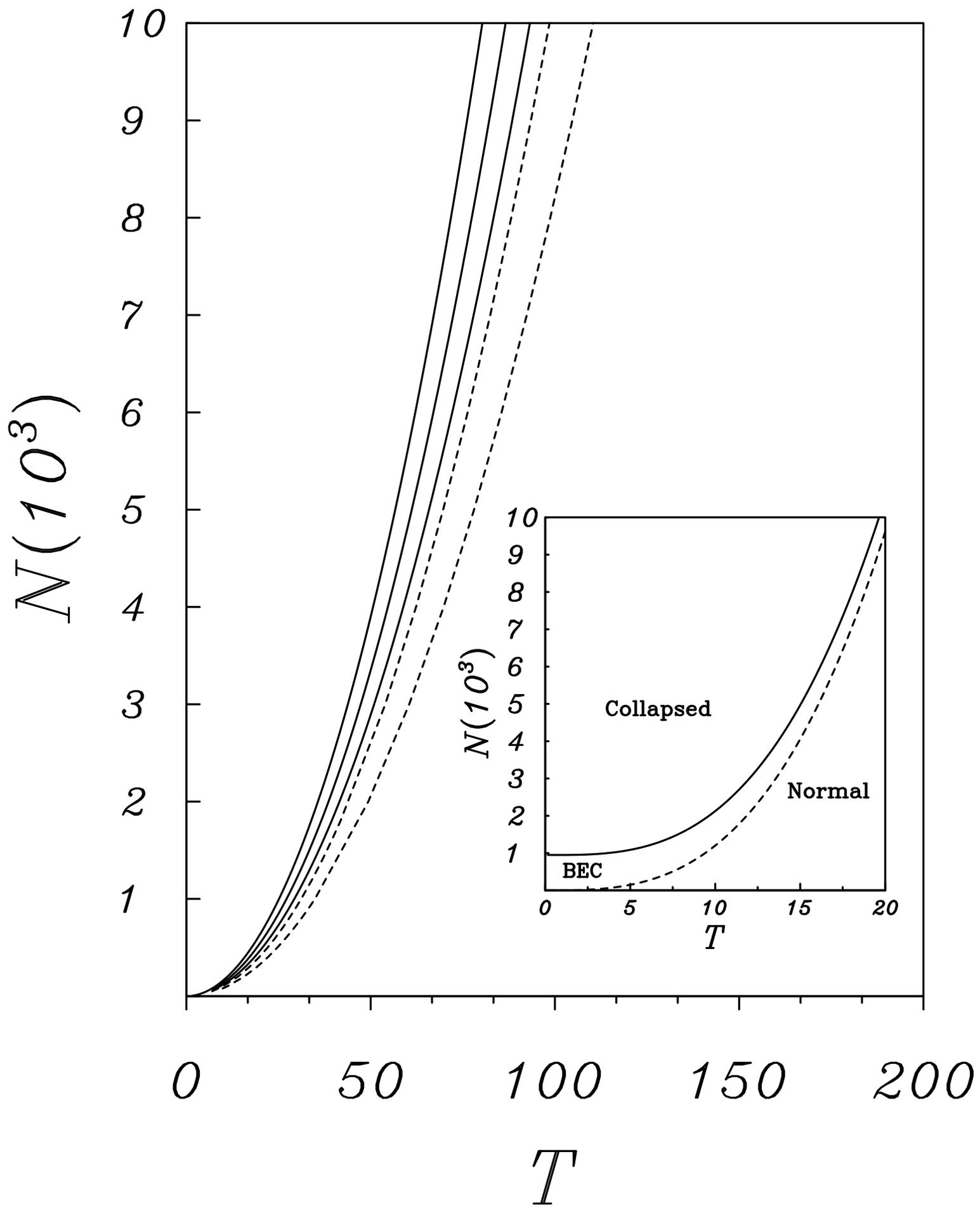
\end{document}